\documentclass[12pt,a4paper]{article}
\usepackage[utf8]{inputenc}
\usepackage{amsmath}
\usepackage{amsfonts}
\usepackage{amssymb}
\usepackage{graphicx}
\usepackage{amsthm}
\usepackage{amssymb}

\newcommand{\bea}{\begin{eqnarray}}
\newcommand{\eea}{\end{eqnarray}}

\title{Oscillation of Dirac and Majorana neutrinos  from muon decay in the case of a general interaction}
\author{Robert Szafron\footnote{robert.szafron@us.edu.pl} \footnote{Present Affiliation: Department of Physics, University of Alberta, Edmonton, AB T6G 2E1, Canada} and Marek Zra\l{}ek\footnote{marek.zralek@us.edu.pl}\\
{\footnotesize Institute of Physics, University of Silesia, }\\{\footnotesize Uniwersytecka 4, PL-40-007 Katowice, Poland}}
\date{August 16, 2012}

\begin{document}
\maketitle
{Keywords: Neutrino oscillation, Muon decay, Non-standard interactions
\\

\abstractname{ We analyse the possibility of distinguishing Dirac and Majorana neutrinos in future neutrino factory experiments in which neutrinos are produced in muon decay when, in addition to a vector type as in the SM, there are also scalar  interactions. We check this possibility in an experiment with a near detector, where the observed neutrinos do not oscillate, and in a far detector, after the neutrino oscillations. Neglecting higher-order corrections, even  neutrino observation in the near detector does not give a chance to differentiate their character. However, this possibility appears in the leading-order after the neutrino oscillations observed in far detector.

}
\section{Introduction}
\

In the Standard Model (SM) with only V--A coupling between neutrinos and charged leptons, massless Dirac and Majorana neutrinos are not distinguishable. In the case of massive neutrinos, whose masses are many orders of magnitude smaller than the masses of charged leptons, the distinction between these two types of neutrino is formally possible but very difficult \cite{Kayser:1982br}. Indeed, the experiments so far have not given us any guidance on the question of whether neutrinos are Dirac or Majorana particles. Probably neutrino-less double beta decay process, if observed, will be the first experiment that could indicate whether the neutrinos are Majorana in nature. When neutrino interactions  have  also non-standard contributions, there are other possibilities for distinguishing between Dirac and Majorana neutrinos. We will discuss one such possibility in this Letter, related to muon decay.  However, the situation is not so simple. In paper \cite{Langacker:1988cm}, muon decay was analysed assuming the most general local, derivative-free, Lorentz invariant and lepton nonconserving interactions. Assuming that neutrinos are not observed and their masses can be neglected, it was shown that up to the leading order in the parameters beyond the SM, it is impossible to distinguish between Dirac and Majorana neutrinos.
In this Letter we will go further and analyse the problem of determining
whether neutrinos are Dirac or Majorana particles when muon decay is used as a source of neutrinos in oscillation experiments (future neutrino factories). We consider neutrino detection in the near (no oscillation) and far (after oscillations) detectors. 

In the next section we will consider Dirac and Majorana neutrinos oscillation in the case of their non-standard interaction. Section 3 contains our conclusions.

\section{Oscillation of Dirac and Majorana neutrinos}
\

We  assume that neutrinos are produced in muon decay as is the case for future neutrino factories \cite{De Rujula:1998hd}. In the SM, muon decay can be described by an effective Lagrangian
\begin{eqnarray}\label{eq:muonSM}
{\cal L}_{I} = -2\sqrt{2}{G_F}\left[ g^V_{i j}  \left(
\overline{e}\gamma^\alpha P_{L}  \nu_i \right)  \left(
\overline{\nu_j}  \gamma_\alpha P_{L}\mu \right) \right]+ h.c.,
\end{eqnarray}
with $g^V$ related to the elements of the neutrino mixing matrix $g^V_{ij} = U_{e i} U^*_{\mu j}$. 
However, if we go beyond the SM, different Lorentz structures can appear.  Present experiments give bounds on the effective couplings for different types of interactions, i.e., scalar vector and tensor, see, e.g., \cite{PDG, delAguila:2009vv}.  A detailed analysis of these constraints shows that from among all possible types of  couplings, the scalar one can give the most significant contributions.
Therefore we confine ourselves to this additional coupling and we take  our Lagrangian to be of the form
\begin{eqnarray}\label{eq:muon}
{\cal L}_{I} = -2\sqrt{2}{G_F}\left[ g^S_{i j} \left(
\overline{e} P_{R}  \nu_i \right) \left(
\overline{\nu_j} P_{L} \mu \right)  + g^V_{i j}  \left(
\overline{e}\gamma^\alpha P_{L}  \nu_i \right)  \left(
\overline{\nu_j}  \gamma_\alpha P_{L}\mu \right)\right]+ h.c.
\end{eqnarray}
Many SM extensions predict additional interactions that contribute to muon decay and generate the effective Lagrangian presented above (see, e.g., \cite{Duka:1999uc, Antusch:2008tz, Medina:2011jh}).  Another reason for choosing this type of Lagrangian is that it is the only type of interaction where for Majorana neutrinos with negligible mass, scalar and SM vector amplitudes interfere even in the limit of the electron mass going to zero\footnote{In subsequent calculations we will neglect the electron mass. Non-vanishing electron mass would introduce another possibility for interference but this effect will be suppressed by powers of the ratio of electron mass to muon mass $\frac{m_e}{M}$ which is below the expected accuracy of future neutrino factories.}, 
giving  the opportunity for leading order effects \cite{Szafron:2009zz}. 
The matrix $g^S$ in Eq.(\ref{eq:muon})  can be chosen freely in agreement with the present bound, while for the vector interaction we want to have it the same value as in the SM so we choose it such that it is proportional to its SM value i.e. $(g^V)^\dagger g^V= || g^V||^2 \hat{P_\mu}$, where $ \hat{P_\mu}$ is a projection operator in neutrino flavour space,  subscript $\mu$ means that it projects on muon neutrino type and $|| g^V||= \sqrt{Tr[(g^V)^\dagger g^V]}$ is the Hilbert-Schmidt norm. In mass base the projection operator $\hat{P_\mu}$ is given by $(\hat{P_\mu})_{ij} =  U_{\mu i} U^*_{\mu j}$. This is equivalent to assuming that in our case 
\bea
g^V_{ij} = g_c U_{e i} U^*_{\mu j},
 \eea
 where $g_c$ is a constant, expected to be close to $1$. A more complicated vector interaction than (3), affecting flavour  symmetry conservation,  has been considered in the literature (see, e.g., \cite{Antusch:2008tz, Coloma:2011rq}), but without any impact on our overall conclusions.
 
It was noticed \cite{Langacker:1988cm} that in muon decay, in the case when the neutrinos are not directly observed and have negligible masses, Dirac and Majorana neutrinos are impossible to distinguish. We will show that this conclusion holds also for the case of a near detector but it is no longer true when neutrinos oscillate.
We will use the density matrix formalism for describing neutrino states, since this properly takes into account the interference between different amplitudes. The usual treatment based on pure states \cite{Grossman:1995wx} cannot be applied here \cite{Ochman:2007vn, Szafron:2010zz, Szafron:2011zz}. The density matrix formalism which we adopt will enable us to describe neutrino states as proper mixed states in the quantum mechanical sense, giving information about the neutrino spectrum produced in muon decay.  We briefly describe this formalism (more detailed information can be found in \cite{Ochman:2007vn, Szafron:2010zz, Szafron:2011zz, Ochman:2010nv}).

 We denote by $A(\lambda, i, E, \omega)$ the amplitude describing  neutrino production in muon decay in the mass state $i$, helicity $\lambda$, and energy $E$; $\omega$ characterises all discrete and continuous degrees of freedom of the other particles that are produced with the neutrinos. Then the matrix elements of the density operator are\footnote{In principle, in our treatment, the density matrix elements should also be labelled by a pair of continuous indices $E, E'$, however using energy--momentum conservation it can be shown that density matrix is diagonal with respect to energy. So we will omit the delta function $\delta(E-E')$ and treat the density matrix as a scalar function of energy.}
 \bea
 (\varrho(E))_{\lambda, i, \lambda' i'}  =\frac{\sum_\omega A(\lambda, i, E, \omega) A^*(\lambda', i', E, \omega)}{N},
  \eea
  where $N$ is a normalisation constant chosen so that 
  $$ \int_{E_{min}}^{E_{max}} dE \sum_{i, \lambda}(\varrho(E))_{\lambda, i, \lambda i}=1,$$
  and our density matrix contains information about the state of the neutrino as well as its energy spectrum. In our calculations we assume that the initial muons are not polarised, so we sum also over the muon polarisation states.
Straightforward but tedious calculations lead to the following result.
The state of a Dirac neutrino produced in muon decay, neglecting the mass of the electron, is 
\bea\label{dm1}
\varrho = \frac{4x^2(2(g^V)^\dagger g^V (3 - 2 x )P_{-}+3(g^S)^\dagger g^S (1-x)P_{+})}{ Tr[(g^S)^\dagger g^S]+4 Tr[(g^V)^\dagger g^V]},
\eea
while in the case of the Dirac antineutrino, we have
\bea\label{dm2}
\varrho = \frac{2x^2(24(g^V)^T (g^V)^* ( 1-x)P_{+}+(g^S)^T (g^S)^* (3- 2 x)P_{-})}{ Tr[(g^S)^\dagger g^S]+4 Tr[(g^V)^\dagger g^V]},
\eea
and for a Majorana neutrino we obtain
\bea\label{mm1}
\varrho = \frac{ K^\dagger K ( 3-2 x)x^2 P_-+6 K^* K^T (1-x)x^2 P_+}{ Tr[K^\dagger K]},
\eea
where the $K$ matrix is given by $K =  g^V+\frac{1}{2}(g^S)^T$, $x=\frac{2 E}{M}$ ($E$ is the energy of the neutrino and $M$ is the muon mass), and $P_+$ (respectively, $P_{-}$) is the positive (respectively, negative) helicity projection operator. The normalisation is such that $\int_0^1 dx Tr[\varrho]=1$.  Since we assume that the initial muons are not polarised, the neutrino density matrices do not depend on the neutrinos' direction of flight.  These formulas (Eqs.(5-7)) are given in the muon rest frame and the lack of manifest Lorentz invariance in our formalism might raise some doubts. Fortunately, for very small neutrino masses, Lorentz transformations do not affect the mass--spin structure of the neutrino density matrix (see, e.g., \cite{Ochman:2007vn}).

For simplicity, in what follows we will assume that in the detection process we measure only the neutrinos with one specific  helicity, e.g., left-handed. This assumption does not restrict the generality of our considerations, since any detection process can be represented by an incoherent mixture of projection operator on both neutrino helicity states\footnote{This basically means that we can measure only left- and right-handed neutrinos and not a linear combination of them.}  In the case when  only one helicity state is detected, the most general  operator describing the detection process is proportional to the projection operator on that one specific helicity. Using the properties of projection operators, we conclude  that all the energy dependence of the density matrix can be factored out in such a case. This energy dependence affects the flux of the neutrinos. Since we are interested in the difference between Dirac and Majorana neutrinos and not the total number of predicted events, this part is not important for our study.  In our case, the factor $( 3-2 x)x^2$ is absorbed into the flux of the neutrinos both for the Dirac and the Majorana case. 
Another important point is that the coupling matrices  $g^V$ and $g^S$ must be normalised in such a way that Eq. (\ref{eq:muon}) leads to a value of the decay rate which is in agreement with precision measurements of this observable. This means that for Dirac neutrinos we must include condition \cite{delAguila:2009vv}
\bea\label{Dircond}
\frac{1}{4}Tr[(g^S)^\dagger g^S]+ Tr[(g^V)^\dagger g^V]=1.
\eea
For the analysis which follows, we can rewrite this condition as
\bea\label{dircond2}
||g^V||=\sqrt{1-\frac{1}{4}||g^S||^2} \approx 1- \frac{||g^S||^2}{8} + O(||g^S||^4).
\eea
 In the Majorana case, the condition is different \cite{Langacker:1988cm}. Due to the interference between the scalar and vector amplitudes,
\bea \label{Majcond}
Tr[K^\dagger K]=1.
\eea
This can be solved for the norm of the matrix $g^V$
\bea \label{majcond2}
||g^V|| &=& 1/2 ( \sqrt{  4 - ||g^S||^2(1 - \alpha^2)}-\alpha ||g^S||)\\
&\approx & 1 - \frac{\alpha ||g^S||}{2} + \frac{1}{8}(\alpha^2-1)||g^S||^2+O(||g^S||^4),
\eea
with
$\alpha =Re(Tr[g^S_N (g^V_N)^*])$, $g^V_N$ being a matrix normalised so that its norm is unity, and similarly for $g^S_N$. We observe that for $\alpha=0$, i.e., no interference between the scalar and vector contributions, this condition is   the same as in the Dirac case.

Adding this up, we can now  consider for the Dirac case a neutrino state described by the matrix
\bea
 \varrho_D = (g^V)^\dagger g^V,
\eea
and for the Majorana case,
\bea\label{roem}
\varrho_M=K^\dagger K,
\eea
with $g^V$ and $g^S$ chosen such that Eqs.(\ref{Dircond} and \ref{Majcond}) are fulfilled. 
The oscillation is described as usual by the unitary transformation
\bea\label{evo}
\varrho_x (L) = U[L] \varrho_x U[L]^\dagger, \;\;\; x = D,M,
\eea
where $U[L]$ sets the neutrino propagation in vacuum or in matter, depending on the experimental conditions.
In general, in the case of vector right-handed interactions, the propagation in matter can also lead to the possibility of distinguishing Dirac and Majorana neutrinos \cite{del Aguila:2007ug}. However we are considering in this Letter only a scalar interaction, so we assume that the propagation Hamiltonian is the same for the Dirac and Majorana neutrinos. In numerical studies, we assume that the neutrinos propagate in vacuum, however the generalisation to matter is straightforward and changes only the quantitative results without changing the general pattern. 

Let us then calculate the number of observed neutrinos. With our assumptions neither the detection cross-section for a typical detection process \cite{Cervera:2000vy} nor the flux depends on the nature  of the neutrino, so we can only consider the oscillation probability
\bea\label{prob}
P_{\mu \rightarrow \alpha}(L)_x = Tr[\varrho_x[L] \hat{P_\alpha}], \;\;\; x = D,M.
\eea 
Here, $\hat{P_\alpha}$ is a projection operator in neutrino flavour space, which projects to  flavour $\alpha$ direction, $(\hat{P_\alpha})_{i,j}=U_{\alpha i} U_{\alpha j}^{*}$.

We neglect the NP contribution to the detection process because it cannot depend on the nature  of the neutrino when no vector right-handed interactions are present.

Calculating the probability factor (\ref{prob}) for $L=0$ using conditions (\ref{dircond2}),  we obtain for the Dirac case 
\bea
P_{\mu \rightarrow \mu}(0)_D& = &1 - \frac{||g^S||^2}{4},\\
P_{\mu \rightarrow e, \tau}(0)_D& =& 0.
\eea
For the Majorana case, using (\ref{majcond2}), we obtain
\bea
P_{\mu \rightarrow \mu}(0)_M& = &1 - \frac{||g^S||^2}{4}(1-\xi_\mu),\\
P_{\mu \rightarrow e, \tau}(0)_M& =& \frac{||g^S||^2}{4}\xi_{e, \tau}.
\eea
 The $\xi_\alpha=Tr[(g^S_N)^* (g^S_N)^T \hat{P_\alpha}]$ depends only on the flavour structure of the matrix $g^S$ and can take any value between $0$ and $1$. The difference between the Dirac and Majorana probabilities  for $L=0$ is of second order in the strength of the scalar interaction: therefore one  expects that this difference is negligible, so we can conclude that \emph{in practice} there is no way  to distinguish Dirac and Majorana neutrinos at the near detector. It is usual, with non-standard interactions, to take into account only effects that are linear in the non-standard parameters and neglect the higher order corrections \cite{Grossman:1995wx}.

The key point in this calculation was the normalisation condition (Eq.(\ref{Majcond})), from which the linear term for the Majorana neutrino disappears, as then the dependence on the New Physics strength is the same for both types of  neutrinos. This no longer occurs when we allow the neutrino to oscillate.  Let us only keep the terms linear in the norm of the scalar interaction. Then a simple calculation for Dirac neutrinos gives ($\beta=e,\mu,\tau$)
\bea
P_{\mu \rightarrow \beta}(L)_D& = &P_{\mu \rightarrow \beta}(L)_{SM} + O(||g^S||^2),
\eea
with $P_{\mu \rightarrow \beta}(L)_{SM}$ the oscillation probability calculated assuming only the SM contribution.
For the Majorana case, from Eqs. (\ref{majcond2}), (\ref{roem}), (\ref{evo}), and (\ref{prob}), we obtain
\bea
P_{\mu \rightarrow \beta}(L)_M =|| g^V||^2P_{\mu \rightarrow \beta}(L)_{SM}+ ||g^V||\cdot || g^S|| \alpha_\beta (L)+ O(||g^S||^2)\\
= P_{\mu \rightarrow \beta}(L)_{SM}-||g^S||(\alpha  P_{\mu \rightarrow \beta}(L)_{SM} -\alpha_\beta(L))   + O(||g^S||^2),\nonumber
\eea
where $\alpha_\beta(L)=Re(Tr[U[L](g^S_N)^\dagger (g^V_N)^T U[L]^\dagger\hat{P_\beta}])$.
 While  obviously the probability factor for Dirac neutrinos still depends quadratically on the strength of the scalar interactions, for the Majorana neutrino the dependence is now linear. It must be kept in mind that the parameters $||g^S||$ have different meanings for Majorana and Dirac neutrinos. This however does not prevent us from concluding that Dirac and Majorana neutrinos can have different oscillation rates. 
 We have also checked this numerically, by generating  random $g^S$ matrices  and calculating the probability factor. The results are given in  Fig. 1,
 which shows the probability $P_{\mu \rightarrow e}(L)$ as a function of $L/E [km/GeV]$ for $||g^S||<0.1$ for both Dirac and Majorana neutrinos. For the Dirac case, the effects of the neutrino scalar interactions are practically indistinguishable from the SM contribution. However for Majorana neutrinos the effects are  larger. 

\begin{figure}\label{fig:1}
\includegraphics[scale=0.5]{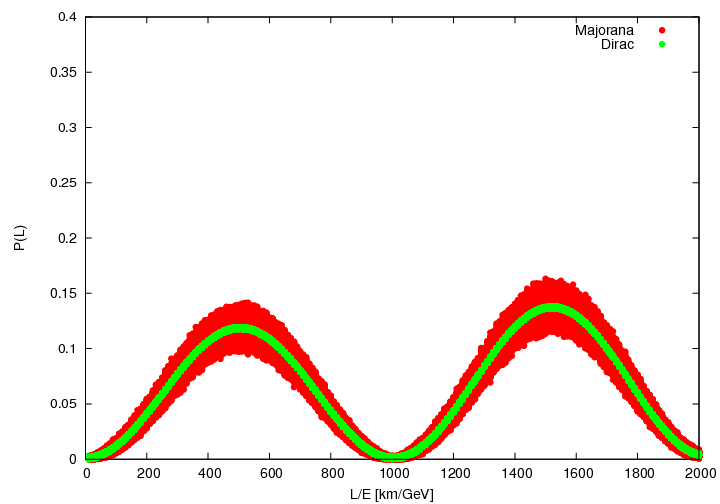}
\caption{$P_{\mu \rightarrow e}(L)$ as a function of $L/E$ for Dirac and Majorana neutrinos. The matrices $g^S$ were randomly generated satisfying $||g^S||<0.1$.}
 \end{figure}
Fig. \ref{fig:2} shows $P_{\mu \rightarrow \mu}(L)$ for $L/E=1000 km/GeV$ as a function of $||g^S||$.  Current experimental limits give $||g^S||<0.55$ \cite{PDG}. From this plot it is clear that observation of the deficit of muon neutrinos on the level outside the allowed range for Dirac neutrinos would indicate that neutrinos are Majorana particles. However, if we observe a number of neutrinos that is in agreement with the current limits for Dirac neutrinos then we cannot say anything about whether neutrinos are Dirac or Majorana particles.
\begin{figure}\label{fig:2}
\includegraphics[scale=0.5]{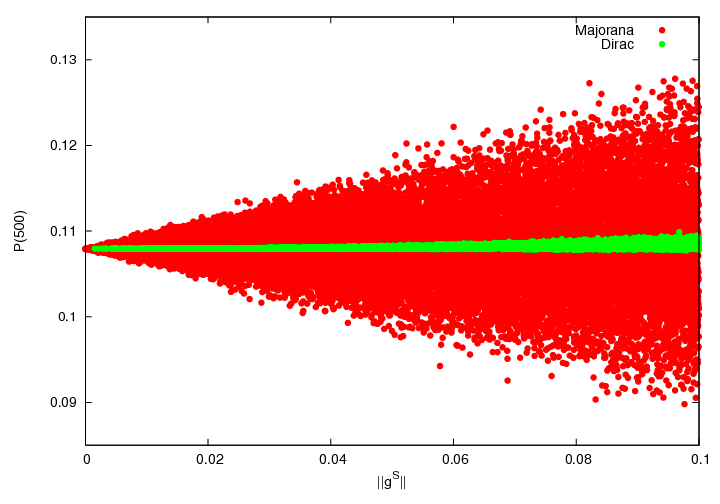}
\caption{$P_{\mu \rightarrow \mu}(L)$ for $\frac{L}{E} =500 \frac{km}{GeV}$ as a function of $|| g^S||$.  }
 \end{figure} 
 
\section{Conclusions}
\

We calculated the oscillation probability for Dirac and Majorana neutrinos in the presence of additional scalar interactions. The dependence on the strength of the new interaction is different for Dirac and Majorana neutrinos. The appearance of a linear term in the expansion for the Majorana neutrinos can in principle help to determine  whether the neutrino is  a Dirac or a Majorana particle. Unfortunately, this analysis is model-dependent and if the deviation were within the allowed range for Dirac neutrinos, then we cannot say anything about the nature of the neutrino. 
When no oscillations are present, i.e., for $L=0$,  in the first order for the transition probabilities the dependence on the strength of the scalar interaction is the same for Dirac and Majorana neutrinos.

Even though these kind of predictions are model dependent and not always conclusive, they can hint at some New Physics and at the determination of the nature of the neutrino as a byproduct of planned experiments, such as a neutrino factory.

\end{document}